\documentclass[12pt]{article}
\usepackage{amsfonts,amsmath,amssymb,epsf}
\usepackage{graphicx,color}
\def\bprt{\bar \partial}
\def\prt{\partial}

\def\d#1{\,{\rm d}#1}

\newcommand{\al}{\alpha'}
\newcommand{\de}{\partial}
\newcommand{\be}{\begin{equation}}
\newcommand{\ba}{\begin{eqnarray}}
\newcommand{\ea}{\end{eqnarray}}
\newcommand{\ee}{\end{equation}}

\newcommand{\f}{\frac}
\newcommand{\s}{\sqrt}

\newcommand{\ti}{\tilde}
\newcommand{\ap}{\alpha}

\newcommand{\ddd}{\cdot\cdot\cdot}
\newcommand{\no}{\nonumber \\}
\newcommand{\la}{\langle}
\newcommand{\lb}{\rangle}
\newcommand{\ep}{\epsilon}

\newcommand{\ket}[1]{{|#1\rangle}}
\newcommand{\bra}[1]{{\langle#1|}}
\newcommand{\Zb}{\mathbb{Z}}

\begin{document}

\begin{titlepage}
\thispagestyle{empty}
\begin{flushright}
hep-th/0303214\\
UT-03-10\\
HUTP-03/A025
\end{flushright}

\bigskip

\begin{center}
\noindent{\large \textbf{Boundary States for D-branes with Traveling Waves
}}\\
\vspace{2cm}
\noindent{
Yasuaki Hikida$^a$\footnote{hikida@hep-th.phys.s.u-tokyo.ac.jp}, \,
Hiromitsu Takayanagi$^a$\footnote{hiro@hep-th.phys.s.u-tokyo.ac.jp} 
and \,
Tadashi Takayanagi$^b$\footnote{takayana@wigner.harvard.edu}}
\\
\vskip 2.5em

{\it $^{a}$ Department of Physics, Faculty of Science,
University of Tokyo\\
Hongo 7-3-1, Bunkyo-ku, Tokyo, 113-0033, Japan\\
\noindent{\smallskip}\\
$^b$ Jefferson Physical Laboratory,
Harvard University\\
Cambridge, MA 01238, USA\\}

\vskip 2em
\end{center}

\begin{abstract}
We construct boundary states for D-branes which carry traveling waves
in the covariant formalism.
We compute their vacuum amplitudes to investigate their interactions.
In non-compact space, the vacuum amplitudes become trivial 
as is common in plane wave geometries.
However, we found that if they are compactified in the 
traveling direction, then the amplitudes are affected by 
non-trivial time dependent effects. 
The interaction between D-branes with waves traveling in the
opposite directions (`pulse-antipulse scattering') are also computed.
Furthermore, we apply these ideas to 
open string tachyon condensation with traveling waves. 
\end{abstract}
\setcounter{footnote}{0}
\end{titlepage}

\newpage

\section{Introduction}
\setcounter{equation}{0}
\hspace{5mm}
String theory on plane wave geometries \cite{plane} has many
interesting features.
In particular, the maximally supersymmetric plane wave solution to type
IIB supergravity is found in \cite{MPP}, and superstrings 
on this background can be exactly solved in the light-cone Green-Schwarz
formalism even in the presence of RR-field \cite{Me}.
This background attracts much attention also because we can discuss its
Yang-Mills theory dual \cite{BMN}.

A plane wave in $D$ dimensional spacetime
is generally defined by the metric (in the Brinkman
coordinate) 
\ba
ds^2=-2dx^+dx^- - \sum_{i,j=1}^{D-2} h_{ij}(x^+)x^ix^j(dx^+)^2
+\sum_{i=1}^{D-2}(dx^i)^2 ,
\ea
and it is time dependent via the term $h_{ij}(x^+)$. 
Physically, this term represents the traveling gravitational waves. 
Furthermore, the background preserves at least a half of supersymmetries.
Thus, it may lead to
a solvable time dependent model with supersymmetry \cite{Tspp}.
(Refer to \cite{HoSt,LMS} for null orbifolds, which have 
 similar properties.) 

However, we have also a disadvantage that we do not know well how
to quantize covariantly string theory in general plane
waves\footnote{In particular, 
superstrings on time {\it in}dependent plane waves (i.e., constant 
$h_{ij}$) with 
NSNS-flux can be described by Nappi-Witten model \cite{NaWi}, 
therefore we can quantize the superstrings covariantly.
Recent developments are given in \cite{NSNS}.
See also \cite{Berkovits}.}. For example,
it is not completely unambiguous 
how to compute even the cosmological constant (i.e., vacuum 
amplitude). 

Motivated by these observations, we would like to
discuss the open string analogue of strings on plane waves. 
In particular, we consider the D-branes with traveling waves in flat 
space; either waves of gauge fields $A^i(x^+,x^i)$ or 
transverse scalar fields $\phi^I(x^+,x^i)$, which are related to
each other by T-duality. 
Indeed, we can show that open string metric on a D-brane with such 
gauge fields leads to a metric of a pp-wave 
(or plane wave if we choose the specific profiles of 
$A^i(x^+,x^i)$) and 
that it is a 1/4 BPS state.
We can choose any functions of $x^+$ as gauge fields or scalar
fields while preserving boundary conformal symmetry as noticed in 
\cite{Callan,BaN,Sk};  
for instance, we can consider a D-brane with a pulse-like world-volume.
Recently, this configuration was examined in the nice paper
\cite{Ba} (see also \cite{DuPi} for more 
generalized models) in the light-cone
gauge of open string, where the world-sheet theory is manifestly 
time dependent. 
There are also some earlier discussions on the related or analogous 
backgrounds, 
see \cite{F1,supercurve} for strings with traveling waves,
and \cite{BaN,My,Ok} for null intersection of D-branes.

It is useful to apply the covariant quantization in order to extract 
information intrinsic to the time dependent physics, and hence
we construct the boundary states representing the D-branes with
traveling waves in the covariant formalism and examine their properties.
These are new type of boundary states in flat spacetime 
with infinite parameters. 
Furthermore, we compute vacuum amplitudes in non-compact and 
compactified 
flat spacetime, where the closed string theory is 
very simple. In the non-compact case
we find a rather trivial result and indeed it is 
the same as that of usual D-brane.
This means that the interaction between the D-branes is the same as the 
usual D-branes, which is consistent with the fact that
there is no vacuum polarization in plane wave background \cite{Gi,De}. 
On the other hand, if we compactify the traveling direction, then we obtain
a very non-trivial amplitude, which reflects the time dependence of
traveling waves. We also compute the interaction between two 
waves (or pulses) traveling in the opposite directions.
It is possible to calculate it only in the covariant formalism. 
We argue that the collision might lead to open string
pair creation like \cite{BaPC} as well as open string tachyonic modes.
Finally, we apply these methods to the open string tachyon
condensation \cite{SeBA}. A configuration with 
traveling open string tachyon is considered, and the corresponding boundary
state is constructed.

The organization of this paper is as follows. In section 2 we 
construct the boundary state for the D-brane with traveling waves and compute
the vacuum amplitude. By using this boundary state, 
we also compute the energy momentum tensor.
In section 3 we consider
the D-branes with traveling waves wrapped on a circle and 
compute the vacuum amplitude. In section 4 we discuss more general 
configuration with traveling waves depending also on the 
other world-volume coordinates. In section 5 we consider the interaction
between two waves traveling in the opposite directions. 
In section 6 we discuss the application to open string tachyon condensation. 
In section 7 we give a brief summary of our results and 
draw conclusions.

\section{Boundary States for D-branes with Traveling Waves}
\hspace{5mm}
There are gauge fields $A^{i}$ and transverse scalar fields $\phi^I$ on
D-branes as massless bosonic fields. The time dependent expectation
values of these fields give interesting time dependent D-brane
backgrounds. Since there is a duality among open string and 
closed string in string theory, this will also lead to an 
intriguing influence on closed strings. We would like to investigate
this issue in simple examples of D-branes with traveling waves. In this 
background, gauge fields or transverse scalar fields depend 
only on $x^+$ and describes waves on D-branes which are traveling at the
speed of light. Interestingly 
the profiles of waves can be chosen arbitrarily as
we explain in the boundary state formalism. Intuitively we can understand
this by the fact that the operators included in 
$A^{i}(X^+)$ and $\phi^I(X^+)$ (or $\ap^+_{n}$) have no 
non-trivial commutation relations with each other \cite{Callan,BaN,Ba}. Below
we construct boundary states for D-branes with gauge field waves 
$A^{i}(x^+)$ and discuss their physical properties. 
The results for D-branes with transverse scalar waves can be 
obtained by T-duality. More general forms of gauge fields will be
considered in section 4.

\subsection{Preparations and Conventions}
\hspace{5mm}
In this paper we define the mode expansion of
closed string in non-compact spacetime as
\ba
X^{\mu}(\tau,\sigma)
 =x^{\mu}+2\al p^{\mu}\tau+i\s{\f{\al}{2}}\sum_{n\neq 0}\f{1}{n}
\Bigl(\ap^{\mu}_{n}e^{-2in(\tau+\sigma)}
+\ti{\ap}^{\mu}_{n}e^{-2in(\tau-\sigma)}\Bigr),\label{modee}
\ea
where the closed string $X^\mu(\tau,\sigma)$ 
has the periodicity under $\sigma \rightarrow \sigma +\pi$. 
The commutation relations are
\begin{equation}
[\alpha^\mu_m,\alpha^\nu_n]=
[\tilde \alpha^\mu_m,\tilde \alpha^\nu_n]=
m\eta^{\mu\nu}\delta_{m,-n},\quad
[x^\mu,p^\nu]=i\eta^{\mu\nu},
\end{equation}
and the other commutators vanish. 
In compactified cases (on a rectangular torus with radii $R^\mu$)
 we should add winding term $2R^{\mu}w_{\mu}\sigma$ to 
the mode expansion (\ref{modee}).

We define the coherent state $|x\lb$ (including only massive modes 
\cite{Ca}), 
which is useful to
construct the boundary state with non-zero gauge flux, as follows 
(here we suppress the index $\mu$ in $\ap^{\mu}_{n}$)
\ba
|x \lb=\prod_{m\geq 1}\exp\left[\f{1}{m}\ap_{-m}\ti{\ap}_{-m}
+\f{x_m}{m}\ap_{-m}-\f{x_{-m}}{m}\ti{\ap}_{-m}
-\f{x_m x_{-m}}{2m}\right] | 0 \lb,\label{x>}
\ea
where we impose $x_m=x_{-m}^{*}$.
This state satisfies
\ba
(\ap_{n}-\ti{\ap}_{-n}-x_{n})|x\lb=0.
\ea
Then the Dirichlet boundary state $|D,\vec{x}\lb$ located at 
$\vec{x}$  
is simply given by $|x=0 \lb\otimes |\vec{x} \lb_{(0)}$.
The state $|\vec{x} \lb_{(0)}$ 
means the zero-mode part of the boundary state and it is normalized 
such that $\la \vec{x} |\vec{x'} \lb_{(0)}=\delta(\vec{x}-\vec{x'})$.
The Neumann boundary state, i.e., D25-brane, is given by the integral
\begin{equation}
\begin{split}
|N\lb&=\frac{T_{25}}{2}\int d\vec{x} \prod_{m\geq 1}\frac{dx_{m}dx_{-m}}
{2\pi m}\ 
|x \lb\otimes |\vec{x} \lb_{(0)}\\&=\frac{T_{25}}{2}
\exp{(-\sum_{m=1}^{\infty}\f{\ap_{-m}\ti{\ap}_{-m}}{m})}
\otimes \int d\vec{x}\ |\vec{x} \lb_{(0)},
\end{split}
\end{equation}
where the normalization constant $T_p$ (for the D$p$-brane) is given by
$T_p=2^{7-p}\pi^{\f{23}{2}-p}
\ap^{'\f{11-p}{2}}$.
The normalization $T_p$ is fixed by making use of
the open-closed duality (Cardy's condition \cite{cardy}). In other
words,
we have\footnote{Here the trace includes the Chan-Paton degrees of freedom
(factor $2$ corresponding to the orientations of open string) 
and zero mode integration as well as the trace over 
string oscillators.}
\ba
\la N|\Delta |N \lb=\int \f{dt}{2t}
\mbox{Tr}[e^{-2\pi tH_{o}}],
\label{Cardy}
\ea
where $H_{o}$ is the open string hamiltonian.
In the compactified case we only have to replace $|\vec{x} \lb_{(0)}$ with
$\sum_{w=-\infty}^{\infty}|\vec{x},w \lb_{(0)}$, where $w$ is the winding
number of the compactified direction.

We should mention that even though in this paper we always use
boundary states in
the (Lorentz) covariant formalism (for a review, see \cite{Di}), 
we suppress the ghost part because it has the usual form
\ba
|\mbox{ghost}\lb=\exp\Big(-\sum_{n=1}^{\infty}(\ti{b}_{-n}c_{-n}
+b_{-n}\ti{c}_{-n})\Big)(c_{0}+\bar{c}_{0})c_1\ti{c}_1|0\lb_{SL(2,R)}.
\ea
The generalizations to the similar boundary states in superstring theory
are also possible in a rather straightforward way (see, e.g., 
\cite{Ca,Di2,Di1,Di}). 
Although in this paper we will mainly show the calculations in bosonic
string theory, most of the results can be easily extended
to the superstring cases (as we will mention later)\footnote{
In the supersymmetric case the normalization is given by
$\frac{T_p}{2}=2^{2-p}\pi^{\f{7}{2}-p}
\ap^{'\f{3-p}{2}}$, which is defined by $|N\lb=\frac{T_p}{2}\f{1+(-1)^F}{2}
|x \lb \otimes | \psi \lb $.}. 
Thus we omit the details of boundary states in superstring theory
for simple expressions.

\subsection{Construction of Boundary States}
\hspace{5mm}
Now we would like to construct the boundary states for D-branes
with traveling waves. For simplicity we will mainly consider the 
spacetime filling D-brane (D25-brane) with gauge fields 
$A_i(X^+)\ \ (i=1,2,\ddd,24)$ in our arguments below. 
The corresponding boundary state, which is denoted as $\ket{P}$,
satisfies the following boundary conditions%
\footnote{Here the gauge field is normalized such that
$B+2\pi\al F$ is the gauge invariant combination.}
\begin{equation}
\begin{split}
&\prt_\tau X^+|P\rangle=0,\\
&(\prt_\tau X^- -2\pi \al \prt_+ A_i(X^+)\prt_\sigma X^i)|P\rangle=0,\\
&(\prt_\tau X^i -2\pi \al \prt_+ A_i(X^+)\prt_\sigma X^+)|P\rangle=0
\quad \mbox{at} \quad \tau=0.
\end{split}
\end{equation}
Naively, we can construct the boundary state $\ket{P}$ 
by multiplying the Neumann boundary state
$\ket{N}$ by the Wilson line, i.e.,
\begin{equation}
\ket{P}=\mbox{P}
\exp\Big(-i\int_0^\pi \d \sigma A_i(X^+)\prt_\sigma X^i\Big)\ket{N},
\label{P>}
\end{equation}
where P denotes the path ordering.

The boundary state must satisfy the boundary conformal invariance
\begin{equation}
(L_n-\tilde L_{-n})|P\rangle=0 \quad \mbox{for}~ {}^{\forall} n,
\label{conformal cond}
\end{equation}
where $L_n$ and $\tilde L_n$ are Virasoro generators on the flat background.
Formally it is easy to show that $\ket{P}$ satisfies 
eq.(\ref{conformal cond}).
However, we should take a great care since the naive expression
(\ref{P>}) would be divergent.
We can avoid the divergence by using the renormalization scheme,
however it breaks boundary conformal symmetry in general.  

Therefore, we should check whether there are divergences or not in the
formal expression (\ref{P>}), and we can easily examine it by using
the boundary state in the path integral formalism \cite{Ca}.
Expanding the gauge field as%
\footnote{We used the periodicity of $A^i$ under $\sigma\to\sigma+\pi$.}
\ba
A^i(X^+)=\sum_{n=-\infty}^{\infty}e^{-2in\sigma}A^i_{n}(X^{+}),\quad
A^i_{n}=A^{i*}_{-n},
\ea
we find the path integral expression and obtain the final
result after the integration
\begin{equation}
\begin{split}
|P\lb &
= \frac{T_{25}}{2}\int d\vec{x} 
\prod_{m\geq 1}\left(\f{1}{2\pi m}\right)^{26}\ dx^\mu_{m}dx^\mu_{-m}
\exp\Bigl[-i\pi\s{2\al}\sum_{n\neq 0}x^{i}_n
A^{i}_{-n}(\hat{x}^+)\Bigr]\\
&\!\!\times \exp\Biggl[\!\!-\f{1}{m}\Bigl(\ap^+_{-m}\ti{\ap}^-_{-m}
\!\!+\ap^-_{-m}\ti{\ap}^+_{-m}\!\!+x^-_m\ap^+_{-m}\!\!-
x^-_{-m}\ti{\ap}^+_{-m}
\!\!+x^+_m\ap^-_{-m}\!\!-x^+_{m}\ti{\ap}^-_{-m} \\ &
\!\!-\f{x^+_{m}x^-_{-m}\!\!+x^-_{m}x^+_{-m}}{2}\Bigr)
\Biggr] \exp\Bigl[\f{1}{m}\ap^i_{-m}\ti{\ap}^i_{-m}
+\f{x^i_m}{m}\ap^i_{-m}-\f{x^i_{-m}}{m}\ti{\ap}^i_{-m}
-\f{x^i_m x^i_{-m}}{2m}\Bigr]| 0\lb
\otimes |\vec{x}\lb_{(0)}, \label{bsp}
\end{split}
\end{equation}
where $A^{i}_{-n}(\hat{x}^+)$ 
is defined by $A^{i}_{-n}(X^+)$ with $X^+(\sigma)$ 
replaced
by 
\begin{equation}
\hat{x}^+(\sigma)=x^++i\s{\f{\al}{2}}\sum_{n\neq 0}
\f{1}{n}x^+_n e^{-2in\sigma}. 
\end{equation}
Since $A^{i}(X^+)$ does not depend on $x^-_{m}$,
we can integrate over $x^-_{m}$ and obtain the delta functions 
$\delta (x^+_m)$.
Therefore we get the finite result  after the integration over $x^+_{m}$ 
(or equally only zero-mode integral $\int dx^{+}_0$).
Then we find that the state (\ref{P>})
satisfies the boundary conformal invariance (\ref{conformal cond})
including renormalization.
We can also understand it in the context of the boundary conformal
field theory (see appendix \ref{ap:BCFT}). It is also straightforward
to construct the similar boundary
state in superstring theory.

\subsection{Energy-Momentum Tensor from Boundary State}
\hspace{5mm}
In this subsection, we compute the energy-momentum tensor and B-field
charge from our boundary state and compare them with the results obtained
in \cite{Ba} as a consistency check.
As discussed in \cite{Sen}
if we expand a boundary state $|B\lb$ for D$p$-brane as
\ba
|B\lb  \propto \int d^{26}k \Bigl[A_{\mu\nu}(k)\ap^{\mu}_{-1}
\ti{\ap}^{\nu}_{-1}+B(k)(b_{-1}\ti{c}_{-1}+\ti{b}_{-1}c_{-1})\Bigr]|k\lb,
\ea
then we can read the value of energy-momentum tensor $T_{\mu\nu}$
as
\ba
T_{\mu\nu}=K(A_{\mu\nu}+\eta_{\mu\nu}B),
\ea
where $K$ is a constant.
The B-field charge $Q_{\mu\nu}$ corresponds to the antisymmetric
part of $A_{\mu\nu}$.

Let us apply this method to our boundary state (\ref{bsp}). 
It is convenient to use the oscillator representation
of $\ket{P}$ obtained by performing integral in eq.(\ref{bsp});
that is
\begin{equation}
\begin{split}
|P\lb
=&\exp\Bigl(\sum_{m=1}^{\infty}(-4\pi^2\al mA^i_{m}(X^+)A^i_{-m}(X^+)
-2\s{2\al}\pi iA^i_{m}(X^+)
\ap^i_{-m}\\
&\ \ \ \ \ \ \ +2\s{2\al}\pi iA^i_{-m}(X^+)\ti{\ap}^i_{-m})
\Bigr)|N\lb.
\label{bspo}
\end{split}
\end{equation}
Expanding $A^i(X^+)$ around 
$x^++2\al p^+\tau$ and dropping $p^+$ by using $p^+|N\rangle=0$, we obtain
\begin{equation}
A^i_m(X^+)|N\rangle=\Big[i\sqrt{\frac{\al}{2}}\prt_+ A^i(x^+)
\frac{(\alpha^+_{m}-\tilde \alpha^+_{-m})}{m}
+ O(\alpha_n^{+2})\Big]|N\rangle.
\end{equation}
Thus we find that the non-zero components are given by 
\begin{equation}
\begin{split}
&B=-\frac{T_{25}}{2},\quad
A_{+-}=\frac{T_{25}}{2},\quad A_{ij}=-\frac{T_{25}}{2}\delta_{ij}
,\quad A_{++}=T_{25}4\pi^2 \ap^{'2}(\de_{+}A^i )^2, \\& \qquad 
A_{+i}=-A_{i+}=T_{25}2\pi\al\de_{+}A^i\ \ \ (i,j=1,2,\ddd,24),
\end{split}
\end{equation}
and hence the energy momentum tensor is
\ba
T_{++}=T_{25}4\pi^2\ap^{'2}(\de_{+}A^i)^2,\ \ 
T_{+-}=T_{25},\ \ 
T_{ij}=-T_{25}\delta_{ij},\ \ Q_{+i}=T_{25}2\pi\al\de_{+}A^i,
\ea
where we set $K=1$ such that the value of energy momentum tensor with
$A^i=0$ agrees with that on the flat background.
Performing T-duality, we can show that the results reproduce the
ones in \cite{Ba} computed by using DBI action.

\subsection{Vacuum Amplitude in Non-compact Space}
\hspace{5mm}
One of the most interesting physical properties we can read from the
boundary states is the interaction between these D-branes. 
This can be computed as the vacuum amplitude, which 
can be directly calculated in our boundary state formalism. 
As we will see below the result turns out to be rather trivial in the
non-compact spacetime.

Let us consider the amplitude between two spacetime filling
D-branes with traveling waves of gauge fields $A^{(1)}_i(x^+)$
and $A^{(2)}_i(x^+)$. We only have to 
evaluate the cylinder amplitude
\ba
Z=\la P^{(2)}|\Delta|P^{(1)} \lb,
\ea
where $|P^{(1)}\lb$ and $|P^{(2)}\lb$ represent the boundary states
 of the form 
(\ref{bsp}) for two different D-branes. We have also defined the 
propagator of closed string as
\ba
\Delta=\f{\al}{2}\int_{0}^{\infty} ds\ e^{-sH_{cl}},
\ea
where $H_{cl}$ denotes the closed string hamiltonian. Then it is easy to 
see that the massive oscillators $\ap^+_{-n}\ \ (n\geq 1)$ included in
$A^i(X^+)$ do not contribute to the amplitude since they cannot
be contracted with $\ap^-_{n}$. This makes the calculation very simple and
indeed we find that the amplitude is the same as that of usual D-brane in
the flat space. To see this, note that there is no zero-mode contribution
to $A^i_{m} \ \ (m\neq 0)$ in (\ref{bspo}).
Almost the same result can also be obtained for
 traveling waves of transverse scalar fields $\phi^{(1)}_i(x^+)$ and
$\phi^{(2)}_i(x^+)$.
The only 
difference is that we have an additional factor $\exp[-\f{2\pi^2\al}{s}
(\phi^{(1)}_i(x^+)-\phi^{(2)}_i(x^+))^2]$, which represents the time dependent
winding energy between the two different D-branes (for details, see 
(\ref{d1p}) in section 3.3). 

These simple results (and their analogous results of closed strings in 
plane wave backgrounds\footnote{In the exactly solvable 
plane wave with NSNS-flux (Nappi-Witten model \cite{NaWi}),
it has been known that the partition function
is the same as that in the flat space \cite{Ki,RTpp}. 
For the plane wave background with RR-field \cite{Me},
a similar result seems to be
difficult to show since there is no solvable 
covariant formalism (see \cite{Ta} for
a  relevant discussion in the operator formalism in the light-cone 
gauge). 
Nevertheless, there are some evidences for the 
triviality of partition function constant \cite{Tsun,Ha}. 
On the other hand, after we compactify a spatial coordinate 
$y(=\frac12 x^+ -x^-)$,
the partition function can be non-trivial as known
in the NSNS plane wave \cite{RTpp} (see also \cite{Su} for 
DLCQ compactification of plane wave). Indeed our results of open string
analogue of plane waves have a similar property as we will show later.})
correspond to the stringy version of 
the known fact that
there are no particle creations and vacuum polarization in 
Yang-Mills or gravitational plane wave background \cite{De,Gi}.
However, things will be different due to winding modes 
if we compactify a spatial coordinate in the light-cone
direction on a circle  
as we will see in the next section.

\section{D-brane with Compactified Traveling Waves}
\setcounter{equation}{0}
\hspace{5mm}
As we have seen above, the vacuum amplitude between D-branes with
traveling waves in flat space turns out to be trivial.
On the other hand, if we study the
open string spectrum in the light-cone gauge \cite{Ba}, then 
we get the non-trivial time dependent world-sheet dynamics. 
This is because the boundary interaction 
$\int d\tau A^i(X^+)\de_{\tau}X_i$ becomes time dependent linear interaction
after we impose the light-cone gauge $X^+=x^+ +2\ap' p^+\tau$ 
with non-zero $p^+$.
Why is there such a difference between these two analyses of 
the same system? The 
answer is that the state with non-zero $p^+$ cannot be 
(easily) expressed in the boundary state formalism
since in the closed string channel the non-zero $p^+$ sector
corresponds to the non-zero winding sector, which does not exist in our 
non-compact space analysis.

In order to see the time dependent effects from the closed string
viewpoint, we compactify the $y$ direction 
(here we assume the coordinates $x^+=t+y$ and $x^-=(t-y)/2$) and study 
the corresponding boundary state.
Mainly we consider the traveling waves of gauge fields on the spacetime
filling brane (D25-brane) in bosonic string theory. 
Later we will also give the results in the transverse scalar case 
and superstring case briefly.
We assume that the gauge field $A_i(X^+)$ obeys the periodicity
\ba
A^i(X^+ +2\pi R)=A^i(X^+),
\ea 
which allows the Fourier expansion ($c_{n}=c_{-n}^*$)
\ba
A^i(X^+)=\sum_{n=-\infty}^{\infty}c^i_{n}e^{-i\f{nX^+}{R}}. \label{gaugef}
\ea

\subsection{Vacuum Amplitude}
\hspace{5mm}
Let us consider the vacuum amplitude between two such spacetime filling 
D-branes.
Since the operators $\ap^{+}_{n}\ (n \neq 0)$ 
commute with each other, we again conclude that only
zero-modes on $A^i_{m}$ in the boundary state (\ref{bspo})
contribute to the vacuum amplitude. By dividing (\ref{gaugef}) into 
zero-modes $x^+,w$ and massive modes $\ap^+_{\pm n}$, we obtain
\ba
A^i(X^+)=
\sum_{n=-\infty}^{\infty}\Bigl(c^i_{n}
e^{-i\f{nx^+}{R}}e^{-2inw\sigma}\Bigr)+\mbox{massive terms}.
\ea

In the non-compact case (i.e., $w=0$), it is easy to see that the zero-mode
term in $A^i_{m}$ is zero
except for $m=0$ as we have seen in the previous section. 
Thus we could effectively regard the boundary state as  
 the usual one
$|N\lb$ in the computation of vacuum amplitude. 

However, in the compactified case an interesting thing does happen.
The zero-mode terms in $A^i_{m}$ are given by
\ba
A^i_{nw}=c^i_{n}e^{-i\f{nx^+}{R}}.
\ea
Thus in the computation of vacuum amplitude we effectively 
obtain a novel form of boundary state, neglecting the massive modes
$\ap^{+}_{n}$ $(n \neq 0)$,
\begin{equation}
\begin{split}
|P\lb\sim&
\exp\Bigl(-4\pi^2\al |w| \sum_{n=1}^{\infty}n|c_{n}|^2
\\& \qquad \qquad +
2\s{2\al}\pi i\sum_{n\geq 1}(c^i_{n}\ap^i_{-n|w|}e^{-i\f{nx^+}{R}\f{w}{|w|}}
-c^i_{-n}\ti{\ap}^i_{-n|w|}e^{i\f{nx^+}{R}\f{w}{|w|}})
\Bigr)|N\lb, \label{bpp}
\end{split}
\end{equation}
which gives a different weight to each winding sector. (Notice that
the sum over $w$ is implicitly hidden in $|N\lb$.) The presence of 
the exponential factor $\sim \exp(-|w|)$ leads to the
suppression for the sectors with large winding number.
This will be interpreted as the 
damping behavior of open string in the time dependent background.
The second term in the exponential
in (\ref{bpp}) represents the linear interaction of open string, which 
is obvious from the form of the boundary interaction.

Now let us compute the amplitude between a D25-brane
with gauge flux $A^{(1)}(x^+)$ and another with $A^{(2)}(x^+)$.
Employing the formula
\begin{equation}
\begin{split}
&\la 0|e^{-\f{1}{m}\ap_{m}\ti{\ap}_{m}}
e^{(f^{(2)}_{-m}\ap_{m}+f^{(2)}_m\ti{\ap}_{m})}
e^{-sH_{c}}
e^{(f^{(1)}_{m}\ap_{-m}+f^{(1)}_{-m}\ti{\ap}_{-m})}
e^{-\f{1}{m}\ap_{-m}\ti{\ap}_{-m}}|0\lb\\
&\ \ =\f{1}{1-e^{-2ms}}
\exp\Bigl(-\f{m}{e^{2ms}-1}\bigl[|f^{(1)}_{m}|^2+|f^{(2)}_{m}|^2
-e^{ms}(f^{(1)}_{m}f^{(2)}_{-m}+f^{(1)}_{-m}f^{(2)}_{m})\bigr]\Bigr)
\label{BHCg},
\end{split}
\end{equation} 
which can be shown by repeatedly applying the Baker-Campbell-Hausdorff's 
formula ($f^{(1,2)}_{m}(=f^{(1,2)*}_{-m})$ 
are arbitrary constants), we obtain\footnote{
Here we have replaced $|w|$ in (\ref{bpp}) with $w$ using 
a symmetry in the final expression of amplitude.}
\begin{equation}
\begin{split}
Z&=\la P^{(2)}|\Delta|P^{(1)} \lb \\
&={\cal N} \int_{0}^{\infty} ds
\sum_{w \in {\mathbb Z}}\exp \Bigl[-\f{w^2R^2s}{2\al}\Bigr]
 \f{e^{2s}}{ \prod_{m=1}^{\infty}(1-e^{-2ms})^{24}}
\exp\Biggl[-\sum_{i=1}^{24}
\sum_{n=1}^{\infty}\f{4\pi^2 \al wn}{e^{2nws}-1} 
 \\ & \qquad \qquad  \qquad \times \Bigl(
(|c^{(1)i}_n|^2+|c^{(2)i}_n|^2)(e^{2nws}+1)-2(c^{(1)i}_n c^{(2)i}_{-n}
+c^{(1)i}_{-n} c^{(2)i}_n) e^{nws}\Bigr)\Biggr]. \label{ampc2}
\end{split}
\end{equation}
The oscillator contributions from $X^+$ and $X^-$ are canceled
by that from the $bc$ ghosts. The normalization
factor ${\cal N}$ is given by ${\cal N}=\f{\al T_{25}^2}{8}V_{26}$ with 
the volume of spacetime $V_{26}=\int dx^+dx^-dx^1\ddd dx^{24}$. 
Note also that the last exponential in (\ref{ampc2})
is always less than one since it can be written as
\begin{equation}
\exp \left[-\sum\f{4\pi^2\al wn}{e^{2nws}-1}
(|e^{nws}c^{(1)i}_{n}-c^{(2)i}_{n}|^2
+|e^{nws}c^{(2)i}_{n}-c^{(1)i}_{n}|^2)\right].
\end{equation}

In particular, if we consider the amplitude between the same D-brane
$c^{(1)i}_{n}=c^{(2)i}_n(\equiv c^{i}_n)$, then 
the non-trivial factor is simplified as
\ba
\exp\Bigl[-8\pi^2 \al w\sum_{i=1}^{24}
\sum_{n=1}^{\infty}n|c^i_n|^2 \Bigl(\f{e^{nws}-1}{e^{nws}+1}\Bigr)\Bigr].
\label{ampcc}
\ea

It is also possible to generalize these results
to the vacuum amplitude between two such D-branes in superstring theory. 
We can employ a similar argument on decoupling of massive modes, and
in conclusion we only have to replace the modular function 
$\f{e^{2s}}{ \prod_{m=1}^{\infty}(1-e^{-2ms})^{24}}
\equiv \eta(is/\pi)^{-24}$ in (\ref{ampc2})
by the familiar terms with theta-functions
\ba
\f{\theta_3(is/\pi)^4-\theta_2(is/\pi)^4-\theta_4(is/\pi)^4}
{2\eta(is/\pi)^{12}}.
\label{theta}
\ea
Therefore the vacuum amplitude vanish due to the supersymmetry. This is 
consistent with the fact that each of the D-branes with traveling waves 
preserve the same eight supersymmetries. 
If one wants to see a similar non-trivial time dependent effect 
even in superstring theory, he or she should consider a brane-antibrane
system. The vacuum amplitude can be obtained by just changing
the sign in front of $\theta_2(is/\pi)^4$ in (\ref{theta}). 
In this case the amplitude does not vanish
and a non-trivial interaction is left.

\subsection{Physical Interpretation of Vacuum Amplitude}
\hspace{5mm}
Now let us consider the physical interpretations of our 
vacuum amplitude (\ref{ampc2}), which represents the interaction between 
two D-branes with traveling waves.
In many familiar examples, 
we can perform modular transformation $s=\pi/t$ and
express the amplitude in the open string channel (see also (\ref{Cardy})).
% obtain open-closed duality (Cardy's condition \cite{cardy}). 
In our case, however, the modular transformation seems very difficult to
perform. Actually, this is natural because we know that the world-sheet
theory is time dependent (for non-zero $p^+$) in the open string side, 
and the open string cylinder amplitude should become complicated. 
Nevertheless, we can extract some information from our amplitude
computed in the boundary state formalism by taking the IR limit $s\to 0$ 
of the open string (on the other hand, $s\to \infty$ 
corresponds to the IR limit of closed string).

First let us consider the IR limit $s\to \infty$ of closed string 
(UV limit $t\to 0$ in open string side).
In this case the important exponential factor becomes
\ba
\sim \exp\Biggl[-4\pi^2 \al |w| \sum_{i=1}^{24}
\sum_{n=1}^{\infty}n
(|c^{(1)i}_n|^2+|c^{(2)i}_n|^2)\Biggr].
\ea 
Thus we have no interaction between the two waves of gauge fields 
$A^{(1)}$ and $A^{(2)}$
since there is no mixing term like $c^{(1)i}_{n}c^{(2)i}_{-n}$.
This is natural because the closed string propagates for a long
distance. 
Note also that the interaction for the winding sectors is suppressed by
the presence of the waves on D-branes.
In the case with a strong pulse 
$\sum_{n=1}^{\infty}n |c_{n}|^2 \gg (\al)^{-1}$, 
no winding mode will propagate between the D-branes. 

On the other hand, in the UV limit $s\to 0$ of closed string (IR limit
$t\to \infty$ of open string), 
the interaction of two waves becomes strong as can be seen from the
non-trivial factor (for $w\neq 0$)
\ba
\sim\exp\Biggl[-4\pi\al t\sum_{i=1}^{24}
\sum_{n=1}^{\infty}|c^{(1)i}_n-c^{(2)i}_n|^2 \Biggr].\label{ampuv}
\ea
This means that the interaction is suppressed for the winding sectors 
except for the case $c^{(1)i}_n=c^{(2)i}_n$ 
(the interaction between the same D-brane).
This is much like a mass shift 
$\delta m^2\sim 2\sum_{n=1}^{\infty}|c^{(1)i}_n-c^{(2)i}_n|^2$ due to
the time dependent Wilson-lines.

Consider the interaction between the same D-brane.
Then, the contribution from the leading order (\ref{ampuv}) vanishes
and the first non-trivial correction to (\ref{ampuv}) is given at
the order $\sim O(w^2s)$.
Combining the first factor in (\ref{ampc2}),
this correction can be regarded as the shift of radius
\ba
R^{'2}=R^2+\f{4}{3}
\pi^2\ap^{'2}\sum_{n=1}^{\infty} n^2\bigl(2|c^{(1)}_n|^2+2|c^{(2)}_n|^2
+c^{(1)}_n c^{(2)}_{-n}+c^{(1)}_{-n} c^{(2)}_{n}\bigr). \label{sra}
\ea
Indeed the shifted radius can be interpreted as the one defined 
by the open string metric $G_{\mu\nu}=g_{\mu\nu}-
(2\pi\al)^2F_{\mu\ap}g^{\ap\beta}F_{\beta\nu}$ 
(see, for example, \cite{SeWi}). 
The corrected radius is estimated as 
\ba
R^{'2}-R^ 2= (2\pi\al R)^2 
\la(F_{+i})^2\lb = 8(\pi\al)^2\sum_{n=1}^\infty n^2|c_n|^2,
\ea
which agrees with (\ref{sra}).
Here we take the average $\la\cdots\lb$ over $x^+$ in the last equality.
% when we consider a single D-brane $c^{(1)}_{n}=c^{(2)}_{n}$,
%in which case the leading term (\ref{ampuv}) vanishes.
After performing the modular transformation,
the mass spectrum of open string
%In this case its mass spectrum
% in the limit $s\to 0$ ($t\to \infty$) 
includes the canonical Kaluza-Klein momentum term $\f{n^2}{R^{'2}}$
for the shifted radius.
%after we employ the modular transformation.

It would also be interesting to ask what will happen if we take
$R\to 0$ limit. For the usual D25-brane (i.e., $c_{n}=0$),
we will obtain a D24-brane by T-duality, under which the winding
mode is identified with the momentum such that $P_{\ti{y}}=wR$.
The momentum $P_{\ti{y}}$ could be finite under the limit $R \to 0$ by
taking $w \to \infty$ with keeping the combination $wR$ finite.
On the other hand, if we naively T-dualize our D25-brane with traveling
waves in the limit, then we get a `exotic' D24-brane with 
only $P_{\ti{y}}=0$ sector in the boundary state since the contributions
from the large $w$ sectors are suppressed as mentioned around eq.(\ref{bpp}).
This means that the D-brane configuration is smeared along the $y$ direction
(this might be natural since the pulse originally runs in the  
$y$ direction). However, in a physical theory, 
only finite energy configurations of gauge fields are allowed. 
This condition is estimated (assuming a weak gauge field) 
as follows\footnote{We may also get stronger conditions from next order 
(${\cal O}(F^4)$) terms.} (for a finite gauge coupling)
\ba
E \propto \int dy \Bigl[(F_{0i})^2+(F_{ij})^2 \Bigr]
\propto \f{1}{R}\sum_{n}n^2|c_{n}|^2.
\ea
Therefore we can see that a D25-brane with a finite energy pulse
becomes an ordinary D24-brane after T-duality transformation.

\subsection{D-strings with Traveling Waves}
\label{D-string}
\hspace{5mm}
As we mentioned above,
the configurations considered are T-dual to D-strings with
waves traveling at the speed of light, thus the boundary
states for these D-strings are obtained by T-dualizing the boundary
states we have constructed.
In order to express the Dirichlet boundary state, it is convenient to use
the following coherent state
\begin{equation}
 \ket{p}= \prod_{m \geq 1} 
   \exp \left[ -\frac{1}{m}\alpha_{-m}\tilde \alpha_{-m} 
    + \frac{p_m}{m}\alpha_{-m} + \frac{p_{-m}}{m} \tilde \alpha_{-m} 
    - \frac{p_{m} p_{-m}}{2m} \right] \ket{0} ,
\end{equation}
which satisfies
\begin{equation}
 (\alpha_m + \tilde \alpha_{-m} -p_m ) \ket{p} = 0 ,
\end{equation} 
where we have imposed $p_m = p^*_{-m}$.
In this basis the Neumann boundary state is given by $\ket{p=0} \otimes
\ket{p = 0}_{(0)}$ with ${}_{(0)}\langle p \ket{p'}_{(0)} = \delta (p-p')$ and
the Dirichlet boundary state $\ket{D1,x^i}$ can be
written as
\begin{equation}
\begin{split}
 \ket{D1,x^i} &= 
% \int dx^{+} dx^- \sum_w \prod_{n \geq 1} \exp 
%  \left[ - \frac{1}{n}(\alpha^+_{-n} \tilde \alpha^-_{-n} 
%         + \alpha^-_{-n} \tilde \alpha^+_{-n} )\right] \ket{x^+,x^-,w}_{(0)}
%     \no & \qquad \times
    \prod_{i=1}^{24} \int dp^i \prod_{m \geq 1}
    dp^i_m dp^i_{-m} 
            \ket{p^i} \otimes e^{-i p^i x^i}\ket{p^i}_{(0)} 
  \otimes \ket{N_{lc}}\\ & 
 = \prod_i \prod_{m \geq 1 } \exp \left( 
     \frac1m \alpha^i_{-m} \tilde \alpha^i_{-m}\right) \ket{0} \otimes
   \int dp^i e^{-i p^i x^i} \ket{p^i}_{(0) } \otimes \ket{N_{lc}} ,
\end{split}
\end{equation}
where $\ket{N_{lc}}$ represents the Neumann boundary state for
light-cone directions. 
This is nothing but the expression of Dirichlet boundary state in
the momentum basis.

By using the basis, we can write the T-dual version of (\ref{bsp}) as
\begin{equation}
\begin{split}
 \ket{P}&= \mbox{P}\prod_i \exp \left( 
   - i \int^{\pi}_0 d \sigma \phi^i (X^+) \partial_{\tau} X^i \right) 
         \ket{D1,0} \\ & =  \prod_i
  \exp \left( \sum_{m=1}^{\infty} (- 4 \pi^2 \alpha' m \phi^i_m \phi^i_{-m} 
              - 2 \sqrt{2 \alpha'} \pi i \phi_m^i \alpha^i_{-m}
              - 2 \sqrt{2 \alpha'} \pi i \phi_{-m}^i \tilde
 \alpha^i_{-m})\right) \\ & \qquad \times
 \exp\left( \sum_n \frac1n \alpha^i_{-n} \tilde
 \alpha^i_{-n}\right) \ket{0} \otimes 
  \int d p^i e^{-  2 \pi i \alpha' p^i \phi^i_0
 (X^+))} \ket{p^i}_{(0)} \otimes \ket{N_{lc}} ,
\end{split}
\end{equation}
where
\begin{equation}
 \phi^i (X^+) = \sum_n \phi^i_n (X^+) e^{-2in\sigma} .
\end{equation}
We should remark that the position of D-string is shifted by the zero
mode $\phi^i_0(X^+)$, which also includes constant shift. 
The amplitude between D-strings with pulses can be calculated as 
\begin{equation}
\begin{split}
Z&=\la P^{(2)}|\Delta|P^{(1)} \lb \\
&={\cal N} \int_{0}^{\infty}
 ds \int dx^+ \left(\frac{2 \pi}{\alpha' s}\right)^{12} 
  e^{-\frac{2 \pi^2 \al}{s} \sum_i
 \left(\phi_0^{(1)i}(x^+) - \phi_0^{(2)i} (x^+)\right)^2}
 \\ & \qquad \times
\sum_{w \in {\mathbb Z}}\exp \Bigl[-\f{w^2R^2s}{2\al}\Bigr]
 \f{e^{2s}}{ \prod_{m=1}^{\infty}(1-e^{-2ms})^{24}} 
 \exp \Biggl[-\sum_{i=1}^{24}
\sum_{n=1}^{\infty}\f{4\pi^2 \al |w|n}{e^{2n|w|s}-1} \\ & \qquad \times
\Bigl(
(|c^{(1)i}_n|^2+|c^{(2)i}_n|^2)(e^{2n|w|s}+1)
 -2(c^{(1)i}_n c^{(2)i}_{-n}
+c^{(1)i}_{-n} c^{(2)i}_n) e^{n|w|s}\Bigr)\Biggr] . 
\label{d1p}
\end{split}
\end{equation}
The non-zero modes of $X^+$ do not contribute to $\phi_0^i$ in the final
expression as before, and hence
the amplitude depends on the distance between two D-branes in the usual
way.

\section{D-brane with More General Gauge Fields}
\hspace{5mm}
In the presence of non-trivial gauge fields, the open string metric can
be written as 
\begin{equation}
 G_{\mu \nu} = \eta_{\mu \nu} - (2 \pi \alpha')^2 
   F_{\mu \rho} \eta^{\rho \sigma} F_{\sigma \nu} ,
\end{equation}
as mentioned above.
We have already dealt with the case of $F_{+i} = h_i(x^+)$, and 
we try to extend our result to the case of more general gauge field in this
section. 
When $F_{+i} = h_{ij}(x^+) x^j$ $(\sum_i h_{ii}=0)$ is included,
the corresponding open string metric becomes
\begin{equation}
ds^2 = -2 dx^+ dx^- -
 \sum_{i,j,k} h_{ij}(x^+) h_{ik}(x^+) x^j x^k (dx^+)^2 + 
   \sum_i (dx^i)^2 ,
\end{equation}
which is the metric of the time dependent plane wave type\footnote{
As argued in \cite{DuPi},  we can include the field strength of the form
$F_{+i}(x^+,x^i)$ with preserving 1/4 supersymmetry, 
and the configuration preserves also the conformal symmetry if
the gauge field satisfies $\de_j \de^j A_+ (x^+ x^i)=0$.
For example, the two cases $F_{+i}=h_i(x^+)$ and $F_{+i} =
h_{ij}(x^+)x^j$ $(\sum_i h_{ii} = 0)$ satisfy the condition.}.

Here we only consider bosonic string theory and D25-brane
with field strength $F_{+i} = 2 h_i (x^+) x^i$ $(\sum_i h_i (x^+) = 0)$ for
simplicity. 
Then, the  boundary state for the D-brane can be written by acting Wilson
line to the Neumann boundary states as
\begin{equation}
\begin{split}
 \ket{P} &=\mbox{P}\exp \left( -i \int^{\pi}_0 
    d \sigma A_+(X^+,X^i)  \partial_{\sigma} X^+ \right) \ket{N}  \\
 & =  \mbox{P} \exp \left( -i \int^{\pi}_0 
    d \sigma \sum_i h_i (X^+) \partial_{\sigma} X^+  X^i X^i  \right) 
    \ket{N} .
\end{split}
\end{equation}
Then the vacuum amplitude\footnote{
Some results of cylinder amplitude in the light-cone gauge 
can be found in \cite{DuPi}.}
 in non-compact space becomes trivial as before.
Thus let us 
again assume $y$ direction is compactified  $X^+ + 2 \pi R \sim
X^+$. 

Since there is a periodicity under $\sigma \to \sigma + \pi$, we can
expand as
\begin{equation}
 h_i (X^+) \partial_{\sigma} X^+ = \sum_{n} H^i_n (X^+) e^{-2 i n \sigma} .  
\end{equation}
By inserting this mode expansion into the previous Wilson line, we can
proceed the calculation as
\begin{equation}
\begin{split}
 \ket{P} &= \int \prod_i dx^i \prod_{m \geq 1} dx^i_m dx^i_{-m} \\
  &\times \exp \left( - i \pi H^i_0 x^ix^i 
   + \sqrt{2 \alpha'}\pi\sum_{n \neq 0}  H^i_{-n} \frac{x^i_n}{n} x^i 
 - i \pi \frac{\alpha'}{2} \sum_{m,n \neq 0} 
   \frac{x^i_m}{m} H^i_{-m+n} \frac{x^i_{-n}}{n} \right) \\
  & \times \exp \left( \frac{\tilde \alpha^i_{-m}  \alpha^i_{-m}}{m}
  + \frac{x^i_m \alpha_{-m}}{m} - \frac{x^i_{-m}{\tilde \alpha^i_{-m}}}{m}
  - \frac{x^i_m x^i_{-m}}{m} \right) \ket{0} \otimes \ket{x^i}_{(0)} 
   \otimes 
  \ket{N_{lc}} ,
\end{split}
\end{equation}
where we use the equality up to normalization.
In this expression, we can perform the Gaussian integral for
$x^i_{m}$ with $m \neq 0$ and obtain 
\begin{align}
 \ket{P} &= \prod_i (\det \Delta^i_{mn})^{-1} \prod_{m,p,n \neq 0} \exp 
 \left( - \frac{1}{2m} a^i_{m} 
 \left( \epsilon(m)\delta_{m,p} - \frac{2\pi i\alpha'}{p}H^i_{-m+p} \right) 
  (\Delta^i)^{-1}_{pn} a^i_{-n} \right) \ket{0} \no
   &\quad \otimes \int dx^i \prod_{p,q \neq 0} 
      \exp \left( - i \pi H^i_0 x^i x^i 
   - 2 \alpha' \pi^2 \frac{H^i_p}{p} (\Delta^i)^{-1}_{pq} H^i_{-q} x^i x^i
 \right. \no
    & \quad \qquad \qquad \quad \left. 
 -  \sqrt{2\alpha'}\pi x^i \left(\frac{H^i_p}{p} (\Delta^i)^{-1}_{pq}
      a^i_{-q} + \frac{a^i_{p}}{p}  (\Delta^i)^{-1}_{pq}
 H^i_{-q} \right) \right) \ket{x^i}_{(0)} \otimes \ket{N_{lc}} ,
\label{ppBS}
\end{align}
We have used  $a^i_{-m} = \alpha^i_m$, $a^i_m = \tilde \alpha^i_{-m}$ for $m
\geq 1$ and defined 
\begin{equation}
 \Delta^i_{mn} = \epsilon(m)\delta_{m,n} 
          + \frac{2 \pi i \alpha'}{n} H^i_{-m+n} ,
\end{equation}
where $\epsilon (m)$ represents the sign of $m$. 
We should notice that
this expression is similar to the boundary state with a constant flux
\cite{AbCa}.
Although it is straightforward to calculate the amplitudes between the
boundary states or closed string states,
it seems that the results cannot be summarized in a simple form.
Thus, we study the amplitudes in a simpler case in the rest of this
section.

When $h_i(X^+) = \mu_i $ $(\sum_i \mu_i = 0)$, equivalently
\begin{equation}
 H^i_0 = 2 R w \mu_i \equiv \frac{w \nu_i}{2 \pi \alpha'} ~, 
 \qquad H^i_n = 0 \quad (n \neq 0 ) ,
\end{equation}
the open string metric becomes that of a time {\it in}dependent plane wave.
In this case, the boundary state (\ref{ppBS}) can be written in a simple
form as
\begin{equation}
\begin{split}
 \ket{\vec \nu}&= \prod_i
  \prod_{m \geq 1}  \left(1 + \frac{i \nu_i w}{m} \right)^{-1}
   \exp \left( - \frac{1}{m}\alpha^i_{-m}
   \left(\frac{m-i \nu_i w}{m+i\nu_i w} 
  \right) \tilde \alpha^i_{-m}\right ) \ket{0} \\ & \qquad \otimes 
   \int dx^i \exp \left(-\frac{i w \nu_i x^i x^i}{2 \alpha'} \right) 
   \ket{x^i}_{(0)} \otimes \ket{N_{lc}} . 
\end{split}
\end{equation}
The amplitude between this type of boundary states is given
by
\begin{equation}
\begin{split}
 Z &= \bra{\vec \nu^{(2)}} \Delta \ket{\vec \nu^{(1)}} \\
   &= \int_{0}^{\infty} ds e^{2s} \sum_{w} 
     \exp \left[ - \frac{w^2 R^2 s}{2 \alpha'}\right] \prod_i
   \left(2 \pi s w^2 \nu_i^{(1)}\nu_i^{(2)}/\alpha' 
  + 2 \pi i w 
 \left(\nu_i^{(1)} - \nu_i^{(2)}\right)/ \alpha' \right)^{-\frac12}
   \\ & \times
  \prod_{m \geq 1} \left(\left(1 + \frac{i \nu_i^{(1)} w}{m}\right)
             \left(1 - \frac{i \nu_i^{(2)} w}{m}\right) -
                         \left(1 - \frac{i \nu_i^{(1)} w}{m}\right)
                         \left(1 + \frac{i \nu_i^{(2)} w}{m}\right)
     e^{-2sm}\right)^{-1} .
%  \no & \times  \prod_{n \geq 1} \frac{1}{(1-e^{-2sn})^{23}} ~.
\end{split}
\end{equation}
It would be interesting
 if we can apply the modular transformation to this
amplitude and interpret it in a open string channel.

\section{Interaction of Pulse and Anti-Pulse}
\begin{figure}[htbp]
  \centerline{\epsfbox{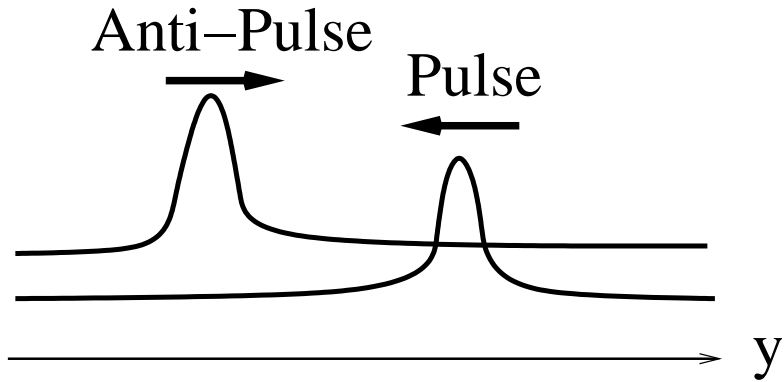}}
	\caption{The collision of pulse and anti-pulse. In this figure
 we consider the T-dualized case, i.e., the pulse-like waves of 
transverse scalars on two D-strings.}
	\label{Fig1}
\end{figure}

\setcounter{equation}{0}
\hspace{5mm}
Next let us proceed to a more complicated and intriguing example, i.e.,
the interaction between two
waves of gauge field $A^i(X^+)$ and $\ti{A}^i(X^-)$
traveling in the opposite directions (see fig.\ref{Fig1}). 
In contrast with the previous examples, we expect non-trivial
particle creation in this example since there is no symmetry in the 
null direction. Thus this configuration leads to a more interesting time
dependent effect.

Since we can use any form of gauge fields $A^i(X^+)$
and $\ti{A}^i(X^-)$, we choose the
ones of pulse-like form. 
In this case we can regard this physical setup 
as a collision between pulse
$P^+$ and anti-pulse $P^-$. 
In superstring theory, $P^+$ and $P^-$ preserve different types of
eight supersymmetries, and hence the presence of both leads to a
non-supersymmetric system. 
 Thus this system is expected to be unstable 
and will tend to decay when the two pulses are approaching.
If the decay occurs completely, it will eventually becomes the
supersymmetric system of
two overlapped D-branes with no gauge flux. However, we cannot
deny the possibility that the annihilation of 
the pulse and anti-pulse takes place partially and smaller pulses remain.

The vacuum amplitude of this system can be computed in our 
boundary state formalism and is non-trivial even in non-compact space 
as we will see below. Another motivation to study this system 
is that the system is the open string analogue of collision of
plane-waves as it is difficult to compute in closed string.
Also it seems impossible to compute it
in open string in the light-cone gauge. 
Thus this example is an interesting application of our boundary state
formulation.

\subsection{D-branes with Constant Null Flux}
\hspace{5mm}
Before we discuss the general system, we would like to examine a `toy
model', i.e., two (spacetime filling) D-branes $|F^+ \lb$ and $|F^- \lb$
with constant gauge flux $F_{+i}$ and $F_{-i}$,
respectively (or $A_i(X^\pm)=F_{\pm i}X^{\pm}$). Essentially 
cylinder amplitude has already been computed in \cite{To} 
by using the analytic continuation of the 
light-cone boundary state (see also \cite{Oh} for open string spectrum and
refer to \cite{Bar} for an earlier literature)
in the context of BPS brane-antibrane system \cite{BaKa}.
Here we will examine it by using the covariant boundary state and clarify
the physical phenomena especially for non-supersymmetric cases. 

Note that we cannot 
reduce this system to well-known case of (purely) magnetic or electric
flux \cite{AbCa,BaPC} by Lorentz transformation.
For simplicity we assume that only
$f_{+}\equiv 2\pi\al F_{+1}$ and $f_{-}\equiv 2\pi\al F_{-1}$ are 
non-zero. Then, the boundary states $|F^\pm \lb$ should 
satisfy the boundary conditions
\begin{equation}
\begin{split}
&(\de_{\tau}X^\mp-f_\pm \de_{\sigma}X^i)|F^\pm \lb=0,\\
&\de_{\tau}X^\pm |F^\pm \lb=0,\\
&(\de_{\tau}X^i-f_\pm \de_{\sigma} X^\pm ) |F^\pm \lb=0.
\end{split}
\end{equation}
These conditions are solved as in (\ref{bsp}) (we only write $|F^+ \lb$)
\begin{align}
|F^+ \lb &=
\exp\Bigl[\sum_{n=1}^{\infty}\f{2}{n}\bigl(
f_+^2\ap^{+}_{-n}\ti{\ap}^{+}_{-n}-f_+\ap^{1}_{-n}\ti{\ap}^{+}_{-n}
+f_+\ap^{+}_{-n}\ti{\ap}^{1}_{-n}\bigr)\Bigr]|N\lb \no
&= \frac{T_{25}}{2}
\prod_{n=1}^{\infty}\f{n}{2\pi}\int d\lambda_n d\lambda_n^*
\exp\Bigl[-\f{n}{2}|\lambda_n|^2-\lambda_n(f_+\ti{\ap}^{+}_{-n}
+\ti{\ap}^{1}_{-n})-\lambda_n^*(f_+ \ap^{+}_{-n}
-\ap^{1}_{-n})\no
&+\frac{1}{n}\bigl(\ap^{+}_{-n}\ti{\ap}^{-}_{-n}+
\ap^{-}_{-n}\ti{\ap}^{+}_{-n}+\ap^{i}_{-n}\ti{\ap}^{i}_{-n}\bigr)\Bigr]
|0\lb \otimes \int d\vec{x}\ |\vec{x}\lb_{(0)}.
\label{BSFL}
\end{align}

Using the second integral expression in (\ref{BSFL}),
we can compute the vacuum amplitude 
$Z_{f}$ between
$|F^+ \lb$ and $|F^- \lb$ as follows;
\begin{equation}
\begin{split}
Z_{f}&={\cal N}\int_{0}^{\infty} ds\ e^{2s}
\prod_{n=1}^{\infty}\f{1}{(1-e^{-2ns})^{24}}\cdot
\left(\f{n}{2\pi}\right)^2\int d\lambda_n d\lambda_n^*
d\mu_n d\mu_n^*\\ 
& \qquad 
\times\exp\Bigl[-\f{n(e^{2ns}+1)}{2(e^{2ns}-1)}(|\lambda_n|^2+|\mu_n|^2)
+\f{ne^{ns}(1-f_+f_-)}{e^{2ns}-1}(\lambda_n\mu_n^*+\mu_n\lambda_n^*)
\Bigr]\\
&={\cal N}\int_{0}^{\infty}
 ds\ e^{2s}\prod_{n=1}^{\infty}\f{1}{(1-e^{-2ns})^{24}}
\cdot\f{(1-e^{-2ns})^2}{(1-2\cos(2\pi\nu) e^{-2ns}+e^{-4ns})},
\end{split}
\end{equation}
where we define 
\ba
\cos(2\pi\nu)=2(f_+f_- -1)^2-1.
\ea
Furthermore, it is possible to write the amplitude
in terms of eta- and theta-functions as
\begin{equation}
Z_{f}=2{\cal N}\sin(\pi\nu)\int_{0}^{\infty} ds\
\f{1}{\eta(\f{is}{\pi})^{21}\ 
\theta_1(\nu|\f{is}{\pi})}.\label{oampf1}
\end{equation}

After performing the modular transformation ($t=\pi/s$) we obtain
\begin{equation}
\begin{split}
Z_{f}&=-2i{\cal N}\sin(\pi\nu)\int_{0}^{\infty} dt\
\f{\pi e^{\pi\nu^2 t}}{t^{13}}
\cdot
\f{1}{\eta(it)^{21}\ \theta_1(-i\nu t|it)}
\\
&=i{\cal N}\sin(\pi\nu)\int_{0}^{\infty} dt\
\f{\pi e^{\pi(2+\nu^2) t}}{t^{13}\sin(\pi i\nu t)}
\prod_{n=1}^\infty
\f{1}{(1-e^{-2\pi nt})^{22}(1-e^{-2\pi nt+2\pi\nu t})
(1-e^{-2\pi nt-2\pi\nu t})}. \label{oampf}
\end{split}
\end{equation}
The supersymmetrization is also easy to be done (recall the Jacobi identity) 
\begin{equation}
\begin{split}
Z_{f}&=\f{{\cal N} \sin(\pi\nu)}{8\sin(\pi\nu/2)^4}\int_{0}^{\infty} ds\ 
\f{\theta_1(\f{\nu+\ep}{2}|\f{is}{\pi})^4}{\eta(\f{is}{\pi})^{8}\ 
\theta_1(\nu|\f{is}{\pi})}\\
&=\f{-i{\cal N\pi\sin(\pi\nu)}}{\sin(\pi\nu/2)^4}\int_{0}^{\infty} dt\ 
\f{e^{-\pi t(\ep^2+2\nu\ep)}}
{8t^{5}}\cdot
\f{\theta_1(-i\f{(\nu+\ep)t}{2}|it)^4}{\eta(it)^{8}\ 
\theta_1(-i\nu t|it)}
 \label{oamps},
\end{split}
\end{equation}
where we should set 
$\ep=0$ for brane-brane amplitude and $\ep=1$ for brane-antibrane one.
We can also check that the above results (\ref{oampf}) and (\ref{oamps}) 
in terms of modular parameter $t$ agree with the open string spectrum
(see \cite{Oh} for open string computations for $f_{+}=f_{-}$ case).

We can see from (\ref{oampf}) and (\ref{oamps}) that there are
infinite number of poles from the factor $\sin(2\pi i\nu t)$
for imaginary values of $\nu$ ($(f_+f_--1)^2>1$). This 
is very similar to the physics on D-branes with (purely) 
electric-field \cite{BaPC}.
Thus, for imaginary $\nu$, the integration over $t$ leads to the
imaginary part of the amplitude, and hence open string pair creations
should occur. 
Notice that $\nu$ takes an imaginary value when  $f_{+}f_{-}<0$.
This is intuitively natural because the electric fields on two D-branes
have the opposite sign. On the other hand, $\nu$ takes a real value for 
$0<f_{+}f_{-}<2$ and the spectrum includes 
open string tachyons induced by the gauge flux. 
Thus open string tachyon condensation should occur in this case.
In the other case $f_{+}f_{-}> 2$, $\nu$ becomes imaginary 
and we observe pair creations.

At the one critical point
$f_{+}f_{-}=0\ (\nu=0)$ the spectrum is the same as the usual D-brane 
since the gauge field $F_{+i}$ (or $F_{-i}$) does not polarize the vacuum
as we have seen above. 
We also have the simplified spectrum at the other critical point
$f_{+}f_{-}=2\ (\nu=1)$. 
The amplitude (\ref{oamps}) between branes ($\ep=0$) is the same as that
of a brane-antibrane without gauge flux, while that between brane and
antibrane ($\ep=1$) is the same as the usual supersymmetric 
amplitude between branes. The latter corresponds to (a generalization
of) the fact found in \cite{BaKa} that a brane-antibrane system with the
critical electric flux and opposite sign of magnetic flux becomes 
supersymmetric.

In summary, the system of two D-branes with constant flux 
$f_{+}$ and $f_{-}$ is unstable in general and should decay via either
open string pair productions or open string tachyon 
condensation\footnote{Recently it was argued in \cite{St} that the
open string tachyon condensation may also lead to another kind of 
open string pair productions.}.

\subsection{Pulse and Anti-Pulse Scattering}
\hspace{5mm}
Now let us turn to the main issue of computing the 
interaction (or equally vacuum amplitude) of pulse and anti-pulse. 
For simplicity we show only the result in bosonic strings. 
The result is not changed substantially even in superstring theory.
The boundary state for a pulse $|P^+ \lb$ is given by
(\ref{bsp}) and that for an anti-pulse $|P^- \lb$ is simply given by 
replacing $\ap_n^+$ in (\ref{bsp}) with $\ap_n^-$.
For anti-pulse, we denote the integration as $dy^\mu_{m}dy^\mu_{-m}$ to
avoid a confusion. Then, the vacuum amplitude is defined as  
\ba
Z_{+-}=\la P_-|\Delta|P_+ \lb, \label{vapa}
\ea
which becomes a rather non-trivial amplitude since we should take
infinitely many contractions of $\ap_n^{+}$ and $\ap^{-}_{-n}$ 
in the Wilson-line terms.
After performing the integration over $x^i_{m}$ and $y^i_{m}$ 
in (\ref{vapa}), 
we obtain 
\begin{equation}
\begin{split}
Z_{+-}
&={\cal N}' \int_{0}^{\infty} ds\ \f{e^{2s}}{\prod_{m=1}^\infty
(1-e^{-2ms})^{24}}\int dx^+dy^-
\prod_{m\geq 1}dx^{\pm}_{m}dx^{\pm}_{-m}dy^\pm_{m}dy^\pm_{-m}\\
&\qquad \times
\exp\Biggl[-\f{4\pi^2\al m(e^{2ms}+1)}{(e^{2ms}-1)}
\Bigl(A^i_{-m}(\hat{x}^+)A^i_{m}(\hat{x}^+)
+\ti{A}^i_{-m}(\hat{y}^-)\ti{A}^i_{m}(\hat{y}^-)\Bigr)\\
& \qquad  \qquad +\f{8\pi^2\al m e^{ms}}{(e^{2ms}-1)}
\Bigl(A^i_{-m}(\hat{x}^+)\ti{A}^i_{m}(\hat{y}^-)
+\ti{A}^i_{-m}(\hat{y}^-)A^i_{m}(\hat{x}^+)\Bigr)\Biggr]\\
&\qquad \times\exp\Bigl[\f{1}{m(e^{2ms}-1)}\Bigl(\f{e^{2ms}+1}{2}
(x^+_m x^-_{-m}+x^-_m x^+_{-m}+y^+_m y^-_{-m}+y^-_m y^+_{-m})\\
& \qquad  \qquad -e^{ms}(x^+_m y^-_{-m}+x^+_{-m} y^-_{m}
+x^-_m y^+_{-m}+x^-_{-m} y^+_{m})\Bigr)\Bigr],
\end{split}
\end{equation}
where the normalization ${\cal N}'$ is defined such that
${\cal N}={\cal N}'\int dx^+ dy^-$.

Finally we integrate out $x^{-}_{m}$ and $y^+_{m}$ as follows;
\begin{align}
Z_{+-}&={\cal N}'\!\! \int_{0}^{\infty} ds\ \f{e^{2s}}{\prod_{m=1}^\infty
(1-e^{-2ms})^{24}}\!\!
\times\!\! \int dx^+ dy^- 
\prod_{m\geq 1}dx^+_{m}dx^+_{-m}dy^-_{m}dy^-_{-m}
\times (e^{ms}-e^{-ms})^2\no
&\times \exp\Bigl[\f{e^{2ms}-1}{4m e^{ms}}
(x^+_m y^-_{-m}+x^+_{-m} y^-_{m})\Bigr]\times
\exp\Biggl[-\f{4\pi^2\al m(e^{2ms}+1)}{(e^{2ms}-1)}
\Bigl(A^i_{-m}(\hat{x}^+)A^i_{m}(\hat{x}^+)\no
&+\ti{A}^i_{-m}(\hat{y}^-)\ti{A}^i_{m}(\hat{y}^-)\Bigr)
\!+\!\f{8\pi^2\al m e^{ms}}{(e^{2ms}-1)}
\Bigl(A^i_{-m}(\hat{x}^+)\ti{A}^i_{m}(\hat{y}^-)\!
+\!\ti{A}^i_{-m}(\hat{y}^-)A^i_{m}(\hat{x}^+)\Bigr)\Biggr]. \label{papt}
\end{align}

Unfortunately it is difficult to perform the integrations in (\ref{papt}). 
Therefore
let us take the slowly changing gauge field limit. Then we obtain
\ba
Z_{+-}\sim{\cal N}'\!\! \int_{0}^{\infty}
 ds\ \f{e^{2s}}{\prod_{m=1}^\infty
(1-e^{-2ms})^{24}}
%\!\!\times\!\! 
\int dx^+ dy^- \de_{+}A_i(x^+)\de_{-}\ti{A}_i(y^-)
\sum_{n=1}^{\infty}\f{16\pi^2\ap^{'2} e^{2ns}}{(e^{2ns}-1)^2}.\no
\label{intapp}
\ea
In the large $s$ limit we can regard this as the closed string exchange
between two D-branes.
For example, the contribution from $m=1$ or $n=1$ part in (\ref{intapp})
represents a massless field 
exchange.

Furthermore, we can also get the full order result with respect to 
$\ap^{'2}\de_{+}A_i(x^+)\de_{-}\ti{A}_i(y^-)$ by neglecting higher
derivatives, and the result is simply given by the previous formula 
(\ref{oampf1}). We should note that the value of $\nu$ depends on
$x^+$, $y^-$ via $F_{+ i}(x^+)$, $F_{- i}(y^-)$ 
and we should also make the integration $\int dx^+dy^-$ explicit. 
Thus, in this approximation, the previous result of toy model 
(with only constant flux) can be utilized. 
Since the pulses, in general, have both positive and negative values of
gauge flux depending on the time and position, 
the collision of pulse and anti-pulse may lead to both open string
creation and open string tachyon condensation. 
These phenomena happen when the pulses approach, and the both effects
should cause the decay of the system at least partially.
A part of the energy may be carried out by the radiations (closed strings).
It is an interesting future problem to find the exact end point of this
unstable system for arbitrary pulses.

\section{Traveling Tachyonic Waves}
\setcounter{equation}{0}
\hspace{5mm}
In a brane-antibrane system, there is an open string 
tachyon field and we can
consider D-brane configuration with a non-trivial tachyon 
profile \cite{SeBA}.
It was shown in \cite{SO(32)} that a non-BPS D-brane can be
described by a tachyonic kink on a brane-antibrane pair by using
conformal field theory and later it was confirmed in \cite{FGLS} by using
boundary state formalism\footnote{A 
tachyon vortex \cite{MaSe} on the brane-antibrane
pair as a marginal deformation 
was discussed in \cite{NTU} by using boundary state description.
See also, e.g., \cite{AsSuTe} for off-shell 
boundary state description of open string tachyon condensation.}.
In the previous analysis, we only include the non-trivial gauge fields
depending on one of the light-cone directions $x^+$.
The configuration with $x^+$ dependent tachyon is also an interesting
system, which we analyze in this section.
As the S-brane \cite{sbrane} 
or rolling tachyon \cite{rolling,Sen} gives an important 
time dependent system in string theory, our 
traveling tachyonic wave (or `null tachyon') may lead to another
one.

Let us consider D9-brane and anti-D9-brane wrapped on a torus with
radii $R$ for $y$ direction and $R^1$ for $x^1$ direction and 
include $\Zb_2$ Wilson line on the anti-D9-brane.
In this configuration the tachyon on a open string stretched between two
branes has an anti-periodic boundary condition and an mode expansion
\begin{equation}
 T ( x^+, x^1) = \sum_n T_{n+1/2} (x^+) e^{i \frac{n+1/2}{R^1} x^1} .
\end{equation}
We have assumed that the tachyon has no dependence of $x^-$
(and also the transverse directions except for $x^1$ direction).
When we restrict the radius to $R^1=\sqrt{\alpha' / 2}$, 
the tachyon of the form
\begin{equation}
 T ( x^+, x^1) = \frac{1}{\sqrt{2}} t (x^+)
 \cos \left( \sqrt{\frac{2}{\alpha'}} (x^1 - Y^1(x^+))\right) 
\end{equation}
becomes an exactly marginal operator for any $t (x^+)$
and $Y^1(x^+)$.

It is known that when $t(x^+)=1/2$ and $Y^1(x^+)=0$, the D9-brane
anti-D9-brane pair with the tachyonic kink is equivalent to a non-BPS
D8-brane \cite{SO(32)}.
The position of the non-BPS D8-brane corresponds to the point of 
$T(x^+,x^1)=0$, thus, in particular, the configuration with $t(x^1)=1/2$
and non-zero $Y^1(x^+)$ describes non-BPS D8-brane at $x^1=Y^1(x^+)$
just like the D-strings in subsection \ref{D-string}.
{}From now on we set $Y^1(x^+)=0$ for simplicity.

The D9-branes on a torus with radii $R$ and $R^1$ 
in the type IIB superstring theory can be described by the boundary states 
\begin{equation}
 \ket{N}_{\rm NS} = \frac{1}{2} [ \ket{N,+}_{\rm NS} - \ket{N,-}_{\rm NS}  ]  
  , \quad
 \ket{N}_{\rm R} = \frac{1}{2} [ \ket{N,+}_{\rm R} + \ket{N,-}_{\rm R}  ] ,   
\end{equation}
and the ghost part. The explicit form in the NSNS sector is given by
\begin{equation}
\begin{split}
 \ket{N,\pm}_{\rm NS} =& \int d \vec x \sum_{w,w^1}
 \exp \left( - \sum_{m \geq 1} 
   \frac{1}{m} \alpha^{\mu}_{-m} g_{\mu \nu} \tilde \alpha^{\nu}_{-m} \right) 
 \\ & \qquad \qquad \qquad \times \exp \left( \pm i  \sum_{r \geq 1/2} 
    \psi^{\mu}_{-r} g_{\mu \nu} \tilde \psi^{\nu}_{-r} \right) 
  \ket{\vec x, w,w^1}_{\rm NS} ,
\label{Npm}
\end{split}
\end{equation}
where $w^1$ is the winding number for $x^1$ direction.
(We concentrate on the NSNS-sector in this section.)
Then, the pair of D9-brane and anti-D9-brane with $\Zb_2$ Wilson line
can be described by the following boundary states
\begin{equation}
 \ket{B,\pm}_{\rm NS} = \ket{N,\pm}_{\rm NS} +  \ket{N',\pm}_{\rm NS} ,~~
 \ket{B,\pm}_{\rm R} = \ket{N,\pm}_{\rm R} -  \ket{N',\pm}_{\rm R} ,
\end{equation}
where $\ket{N',\pm}$ includes the $\Zb_2$ Wilson line.

When $R^1 = \sqrt{\al/2}$, we can fermionize the boson 
$X^1 (\tau,\sigma) = \frac12 (X^1 (\tau + \sigma) + \tilde X^1 (\tau -
\sigma))$ as
\begin{equation}
\begin{split}
 e^{ \pm \frac{i}{\sqrt{2 \al}} X^1 (\tau + \sigma)} &\simeq
 \frac{1}{\sqrt2} ( \eta (\tau + \sigma) \pm i \xi (\tau + \sigma) )
 , \\
 e^ {\pm \frac{i}{\sqrt{2 \al}} \tilde X^1 (\tau - \sigma) } &\simeq
 \frac{1}{\sqrt2} ( \tilde \eta (\tau - \sigma) 
   \pm i \tilde \xi (\tau - \sigma) ) .
\end{split}
\end{equation}
By using the fermionic partners $\psi^1$ and $\tilde \psi^1$, 
we can define a new bosonic field  $\phi (\tau, \sigma) = \frac{1}{2}(
\phi (\tau + \sigma) + \tilde\phi(\tau -\sigma))$ by
\begin{equation}
\begin{split}
 \frac{1}{\sqrt2} ( \xi (\tau + \sigma) \pm i \psi^1 (\tau + \sigma) ) &\simeq
 e^{ \pm \frac{i}{\sqrt{2 \al}} \phi (\tau + \sigma)}
 , \\
 \frac{1}{\sqrt2} ( \tilde \xi (\tau - \sigma) 
   \pm i \tilde \psi^1 (\tau - \sigma) ) &\simeq
 e^ {\pm \frac{i}{\sqrt{2 \al}} \tilde \phi (\tau - \sigma) } .
\end{split}
\end{equation}
The advantage of the redefinition of coordinates is that we can rewrite
the tachyon vertex in a simple form (in the zero picture) as \cite{SO(32)}
\begin{equation}
 V_T =
 \frac{i t (X^+)}{\sqrt{2 \al}} \partial_{\sigma} \phi (\sigma) 
   \otimes \sigma^1 .
\label{tachyon}
\end{equation}
The sigma matrix $\sigma^1$ corresponds to the Chan-Paton factor.

In the new coordinate system with $\phi$ and $\eta$ ($\tilde \eta$),
the Neumann boundary state can be constructed by replacing $X^1$ and
$\psi^1$ by $\phi$ and $\eta$ \cite{FGLS,NTU} as
\begin{align}
 \ket{B,\pm}_{\rm NS} &= \sum_{w,w^{\phi}} 
 \int \prod_{\rho,\mu,\nu \neq 1} d x^{\rho} d x^{\phi} 
 \exp \left( - \sum_{m \geq 1} 
   \frac{1}{m} \alpha^{\mu}_{-m} g_{\mu \nu} \tilde \alpha^{\nu}_{-m} \right) 
 \exp \left( \pm i  \sum_{r \geq 1/2} 
    \psi^{\mu}_{-r} g_{\mu \nu} \tilde \psi^{\nu}_{-r} \right) 
 \no & \quad \times 
 \exp \left( - \sum_{m \geq 1} 
   \frac{1}{m} \alpha^{\phi}_{-m}\tilde \alpha^{\phi}_{-m} \right) 
 \exp \left( \pm i  \sum_{r \geq 1/2} 
    \eta_{-r}  \tilde \eta_{-r} \right) 
  \ket{x^{\rho}, x^{\phi},  w, 2w^{\phi}}_{\rm NS} .
\end{align}
The mode expansions of $\phi$ and $\eta$ are given in the way similar to
$X^1$ and $\psi^1$.
Using the tachyon vertex operator (\ref{tachyon}), 
we obtain the boundary state
for D9-brane anti-D9-brane pair with the tachyonic kink as 
\begin{align}
 \ket{T,\pm}_{\rm NS} &= \frac12 { \rm PTr}
  \exp \left( i \int^{\pi}_0 d \sigma
   \frac{ t (X^+)}{\sqrt{2 \al}} \partial_{\sigma} \phi (\sigma) 
   \otimes \sigma^1 \right) \ket{B,\pm}_{\rm NS} \no
&=  \frac12 \exp \Bigl(- 2 \pi^2 \sum_{m=1}^{\infty} m t_{m}t_{-m} \Bigr) 
   \Bigl[\exp \left(2 \pi i (t_{m}\ap^{\phi}_{-m}
 - t_{-m}\ti{\ap}^{\phi}_{-m} +  t_0 w^{\phi}) \right) 
 \no & \qquad \qquad +
   \exp \left(- 2 \pi i (t_{m}\ap^{\phi}_{-m}
 - t_{-m}\ti{\ap}^{\phi}_{-m} +  t_0 w^{\phi} \right)  \Bigr]
\ket{B,\pm}_{\rm NS} ,
\end{align}
where we defined the mode expansion as $t(X^+)=\sum_{n}t_n e^{-2in\sigma}$.
Therefore, we conclude that the configuration with a traveling open
string tachyon can be analyzed in the way similar to the ones with
traveling gauge fields. For example, 
the amplitudes between these boundary states can be calculated
as we have done in (\ref{ampc2}).

\section{Summary and Conclusions}
\hspace{5mm}
In this paper we have investigated several properties of D-branes
with traveling waves in the covariant boundary state formalism.
The traveling waves are carried by gauge fields or transverse 
scaler fields.
Interestingly, the boundary states have a novel feature that 
they have infinitely many parameters, which describe every forms of  
traveling waves.

Employing the boundary states and computing their vacuum amplitudes,
we analyzed the interactions between these D-branes. We found that 
in non-compact spacetime the interactions are the same as those
between usual D-branes.
 However, in the compactified case the interactions turn out to be
very non-trivial and they depend on the form of the waves explicitly.
The non-trivial contribution comes from winding modes of closed strings 
in the compact space, as we can see it directly in the boundary state. 
In this way we found that the time dependence affects
the interaction between D-branes with traveling waves 
if we compactify the space in the traveling direction.

We also generalized the form of waves such that the waves depend on the 
spatial coordinate other than $x^+$, and obtained the vacuum
amplitude in a simplified case. In the example,
the open string metric on the D-brane becomes that of
a plane-wave not of a general pp-wave.

By using our formalism it is also possible to compute 
the interaction between two D-branes with waves traveling in the
opposite direction; the configuration 
seems difficult to analyze in the light-cone gauge. 
This leads to an interesting non-supersymmetric time dependent system 
in superstring theory. 
In particular, we can choose the form of wave such that it represents
the collision of pulse and anti-pulse.  
We obtained the integral formula of the vacuum amplitude and 
argued that the open string creation or tachyon condensation may occur 
and it may lead to a (partial) decay of this unstable system.

Finally we considered the application of these traveling wave 
configurations to the open string tachyon condensation. 
We obtained a new boundary state which represents the traveling 
tachyonic waves.

\subsection*{Acknowledgments}
\hspace{5mm}
We are grateful to T. Harmark, S. Minwalla, S.-J. Rey,  Y. Sugawara, 
F. Sugino and N. Toumbas for useful discussions. We also thank B. Durin and
B. Pioline for e-mail correspondence.
The work of TT was in part supported by DOE grant DE-FG03-91ER40654. 

\vskip2mm

\appendix

\section{Pulse D-brane as Boundary Deformation}
\label{ap:BCFT}
\setcounter{equation}{0}
\hspace{5mm}
In this appendix we would like to discuss D-branes with traveling
waves from the point of view of the boundary deformation \cite{ReSc}.
When we consider the (normal) Neumann boundary condition,
the gluing condition is given by
\begin{equation}
\prt X^\mu(z)=\bprt X^\mu(\bar z)\quad \mbox{at}\quad z=\bar z,
\label{normal Neumann}
\end{equation}
where the world-sheet is the upper half plane of $z$ and
the boundary is Im$(z)=0$.
The OPE between bulk fields is given by
\begin{equation}
X^\mu(z_1,\bar z_1)X^\nu(z_2,\bar z_2)=
-\frac{\al}{2}\eta^{\mu\nu}\Big[\ln |z_1-z_2|^2
+\ln |z_1-\bar z_2|^2\Big]+\mbox{reg},\label{Neumann OPE}
\end{equation}
and we also obtain the OPE between boundary fields by setting
$z_i=\bar z_i=x_i,$
\begin{equation}
X^\mu(x_1)X^\nu(x_2)= -2\al \eta^{\mu\nu}\ln |x_1-x_2| +\mbox{reg}.
\label{BOPE}
\end{equation}

Next let us deform the gluing condition
(\ref{normal Neumann}) to that with traveling waves
\begin{equation}
\prt_\tau X^i-2\pi \al \prt_\sigma A^i(X^+)=0
\quad \mbox{at}\quad \tau=0,
\label{null Neumann}
\end{equation}
where relations between $(\sigma,\tau)$ and $(z,\bar z)$
in this appendix are given by
\begin{equation}
z=(\sigma+\tau),\quad \bar z=(\sigma-\tau),\quad
\prt=\frac{1}{2}(\prt_\sigma+\prt_\tau),\quad
\bprt=\frac{1}{2}(\prt_\sigma-\prt_\tau).
\end{equation}
As we will find, the deformation is generated
by the boundary marginal field ${\cal A}$
\begin{equation}
{\cal A}(\sigma)
\equiv -iA_i(X^+(\sigma))\prt X^i(\sigma)=
i\int \d k^{-} c_i(k^-)
e^{-ik^-X^+(\sigma)}\prt X^i(\sigma), \label{deform op}
\end{equation}
where the Fourier mode $c_i(k^-)$ satisfies $c^\ast_i(k^-)=c(-k^-)$.
Notice that the operator ${\cal A}$ is self-local (in the terminology of
\cite{ReSc})
because the OPE between ${\cal A}$'s is given by
\begin{equation}
{\cal A}(x_1){\cal A}(x_2)=  
\frac{2\al A_i(x_1)A^i(x_2)}{(x_1-x_2)^2}+\mbox{reg}.
\end{equation}

\subsection{Boundary Conformal Invariance}
\hspace{5mm}
In order to preserve boundary conformal symmetry
under the deformation, it is necessary that there is
no mixing between the deformation
field ${\cal A}$ and any marginal field. Thus it is needed that 
the following three point functions for all marginal fields $\psi$
vanish 
\begin{equation}
\langle {\cal A}(x_1){\cal A}(x_2)\psi(x_3)\rangle
=0\label{mixing}.
\end{equation}
By using the momentum conservation and
the contraction of $\prt X^i$, we can trivially show
that eq.(\ref{mixing}) is satisfied for all marginal fields
except $\psi=\prt X^-$. Thus we only need to
calculate the boundary three point function
\begin{equation}
\langle {\cal A}(x_1){\cal A}(x_2)\prt X^-(x_3)\rangle
\equiv \frac{C_{{\cal A}{\cal A}P^-}}{(x_1-x_2)(x_2-x_3)(x_3-x_1)}.
\end{equation}
Using the correlation function
\begin{equation}
\begin{split}
&\langle :e^{-ik_1^-X^+(x_1)}: :e^{-ik_2^-X^+(x_2)}: 
\prt X^i(x_1) \prt X^j(x_2)\prt X^-(x_3)
\rangle_N\\
& \qquad \qquad \qquad = iC \delta (k_1^- + k_2^-)\Big[
2i \al \Big(\frac{k^-_1}{x_3-x_1}+\frac{k^-_2}{x_3-x_2}\Big)
\Big]\frac{(-2\al \delta^{ij})}{(x_1-x_2)^2}\\
&\qquad \qquad \qquad =-\frac{4\alpha^{\prime 2}C\delta^{ij}
\delta(k_1^- + k_2^-) k^-_1}
{(x_1-x_2)(x_2-x_3)(x_3-x_1)},
\end{split}
\end{equation}
with a constant $C$, we obtain
\begin{equation}
\begin{split}
C_{{\cal A}{\cal A}P^-}&= 4\alpha^{\prime 2}
\int \d k_1^{-} \d k_2^- 
c_i(k^-_1)c_i(k^-_2)
\delta(k_1^- + k_2^-) k^-_1\\
&= \frac{2\alpha^{\prime 2}}{\pi}
\int \d y \int \d k_1^{-} \d k^-_2
c_i(k^-_1)c_i(k^-_2)k^-_1e^{-iy(k^-_1+k^-_2)}\\
&=\frac{2i\alpha^{\prime 2}}{\pi}\int \d y
\frac{d}{d y}A^2_i(y)=0.
\end{split}
\end{equation}
Thus we conclude that eq.(\ref{mixing}) is satisfied for all marginal
fields. 

\subsection{Boundary Deformation}
\hspace{5mm}
Finally we show that the marginal field ${\cal A}$
deforms the (normal) Neumann condition to
that with traveling waves (\ref{null Neumann}).
The gluing
condition is deformed by the deformation as follows \cite{ReSc}
\begin{equation}
0=\lim_{\delta \rightarrow +0}
\sum_{n=0}^\infty
\frac{1}{n!}\int_{\gamma_1}\cdots \int_{\gamma_n}
\d \sigma_1 \cdots \d \sigma_n {\cal A}(\sigma_1)
\cdots {\cal A}(\sigma_n)
[\prt X^i(z_{\delta})-\bprt X^i(\bar z_\delta)],
\label{def of deform}
\end{equation}
where we define $z_\delta =x+2i\delta$, $\bar z_\delta=x-2i\delta$
and $\gamma_p$ as the line Im$(\sigma_p)=\frac{\epsilon}{p}$
($\epsilon < \delta$).
Notice that we have regularized eq.(\ref{def of deform}) by
analytically continuing ${\cal A}(\sigma)$ into the
upper half plane.
Since the self local operator ${\cal A}$ has the important property
\begin{equation}
\int_{\gamma_1} \d z \int_{\gamma_2} \d z' {\cal A}(z)
{\cal A}(z')=0,\label{locality}
\end{equation}
where we assume that there is no insertion in Re$(z)> 0$,
there is no contribution from the $n > 1$ part of (\ref{def of deform}).
Thus we have to examine only the $n=1$ part, which is calculated as
\begin{equation}
\begin{split}
\int_{-\infty}^\infty
\d \sigma {\cal A}(\sigma)\prt X^j(z_2)
=-2\pi \al \prt A^i(\tilde X^+(z_2)),\quad
\int_{-\infty}^\infty
\d \sigma {\cal A}(\sigma)\bprt X^j(\bar z_2)=0,
\end{split}\label{hol n1}
\end{equation}
where we defined $\tilde X^+(z)\equiv \frac12 (X^+_L(z)+X^+_R(\bar z))$.
Therefore the deformed boundary condition is obtained as follows
\begin{equation}
\begin{split}
0&=\lim_{\delta \rightarrow +0}
\Big[1+\int_{-\infty}^\infty \d \sigma {\cal A}(\sigma)\Big]
[\prt X^i(z_\delta)-\bprt X^i(\bar z_\delta)]\\
&=\prt_\tau X^i(x)-2\pi \al 
\prt_+ A^i(X^+(x)) \prt_\sigma X^+(x).
\end{split}
\end{equation}
This is nothing but the condition (\ref{null Neumann}).

\end{document}